\documentclass[14pt, a4paper]{article}

\usepackage{arxiv}
\usepackage{amsmath}
\usepackage{amssymb}
\usepackage{amsfonts}
\usepackage{graphicx}
\usepackage{tikz}
\usepackage{circuitikz}
\usepackage{pgfplots}
\usepackage{geometry}
\usepackage{booktabs}
\usepackage{setspace}
\usepackage{caption}
\usepackage{subcaption}

\pgfplotsset{compat=1.18}

\title{Electromagnetic Characterization of Magnetic Ring: Case of Circular Cross-Section Shape}

\author{%
  Taha El Hajji*,\quad Lars Sjöberg \\
  Alvier Mechatronics AB, Helsingborg, Sweden  \\
  \texttt{taha.elhajji@gmail.com}
}

\begin{document}

\maketitle

\begin{spacing}{1.8}
\begin{abstract}
This paper introduces a comprehensive two-dimensional analytical model of a toroidal magnetic ring with circular cross-section under sinusoidal excitation. Applying Maxwell’s equations in local polar coordinates within a complex permeability, the model derives analytical expressions for the internal magnetic field, magnetic flux, complex impedance, and total losses. It rigorously separates the contributions of eddy current losses, hysteresis losses, and winding losses, while explicitly incorporating the skin effect in the conductive core via Bessel functions. An expression for the apparent permeability is also provided, enabling the nonlinear core behavior to be mapped onto simplified linear material models. The resulting analytical model offers a computationally efficient and accurate foundation for standardized magnetic material characterization, such as Brockhaus and Iwatsu ring measurements, as a powerful alternative to 2D and 3D finite element analysis.
\end{abstract}
\end{spacing}

\newpage

\section{Introduction}
The development of high-performance electromagnetic systems heavily depends on the precise characterization of soft magnetic materials. Standardized material evaluation methods, utilizing equipment like Brockhaus or Iwatsu analyzers, typically involve winding a toroidal specimen to measure its current-voltage characteristics. However, the transition toward high-frequency power conversion, spanning from conventional 50\,Hz applications up into the MHz regime, exposes fundamental limitations in these traditional measurement frameworks. 

The primary challenge arises from the invalidation of the uniform magnetic flux assumption at elevated frequencies. Within conductive magnetic cores, induced eddy currents generate a pronounced skin effect, driving the magnetic flux radially outward toward the surface. For toroidal rings with a circular cross-section, this phenomenon manifests as a highly non-uniform, radially dependent internal field that conventional empirical models cannot accurately reproduce \cite{ieee4527040}. Coupled with inherent material hysteresis, which creates a phase shift between the applied magnetic field and the induced flux, it becomes essential to adopt a complex permeability framework to properly characterize these frequency-dependent energy storage and dissipation mechanisms \cite{sd_ferrite}.

In the past, engineers have predominantly relied on purely empirical formulations, such as the Steinmetz equation \cite{ieee9003248}. While useful for basic approximations, these models fail to capture the physical realities of internal spatial field distributions or the magnetic shielding effects dictated by the core geometry. Conversely, while Finite Element Analysis (FEA) provides deep insights into these local distributions, its high computational time makes it impractical for rapid and iterative material characterization. Consequently, there is a renewed research focus on formulating exact analytical solutions for specific core geometries to efficiently evaluate eddy current dynamics and macroscopic volume losses \cite{ieee668058, ieee5382736}.

Addressing this gap, the present work develops a rigorous 2D analytical model of the complex impedance for a toroidal ring with a circular cross-section, explicitly accounting for the coupled effects of hysteresis and eddy currents. A similar mathematical model addressing the case of a square cross-section toroid was also developed by authors \cite{TahaArxivRingS}. By solving Maxwell's equations within a local polar coordinate system, this paper derives exact, closed-form expressions for the internal fields, magnetic flux, overall impedance, and core losses. The integration of complex permeability enables a precise mathematical uncoupling of hysteresis and eddy current losses, relevant for accurate core loss modeling without the need for extensive FEA simulations.

\section{Nomenclature}
\begin{itemize}
    \item $D_m$: Mean diameter of the toroidal ring [m]
    \item $ID, OD$: Inner and Outer diameter of the ring [m]
    \item $a$: Radius of the circular cross-section, $a = (OD - ID)/4$ [m]
    \item $A$: Cross-sectional area, $A = \pi a^2$ [m$^2$]
    \item $l_m$: Mean magnetic path length, $l_m = \pi D_m$ [m]
    \item $\mu_c$: Complex magnetic permeability, $\mu_c = \mu' - j\mu'' = |\mu_c| e^{-j\theta}$ [H/m]
    \item $\mu', \mu''$: Real and Imaginary parts of complex permeability [H/m]
    \item $\theta$: Hysteresis loss angle [rad]
    \item $\sigma$: Electrical conductivity of the core material [S/m]
    \item $\omega$: Angular frequency, $\omega = 2\pi f$ [rad/s]
    \item $N_1, N_2$: Number of turns of primary and secondary windings
    \item $I$: RMS current supplied to the primary [A]
    \item $V_1, V_2$: Voltage across windings [V]
    \item $E_1$: Induced electromotive force (EMF) [V]
    \item $\mathbf{H}, \mathbf{B}, \mathbf{E}, \mathbf{J}$: Standard electromagnetic field vectors
    \item $R_{DC}$: DC resistance of the primary winding [$\Omega$]
    \item $R_{winding}$: AC resistance of the winding accounting for skin effect [$\Omega$]
    \item $Z_{core}$: Complex impedance of the core [$\Omega$]
    \item $Z$: Total complex impedance [$\Omega$]
    \item $Y_{core}$: Complex admittance, $Y_{core} = G_{core} - jB_{core}$ [S]
    \item $G_e, G_h$: Eddy current and Hysteresis equivalent conductances [S]
    \item $P_e, P_h, P_w$: Eddy current, Hysteresis, and Winding power losses [W]
    \item $k$: Complex wave number in the core, $k = \sqrt{-j\omega \mu_c \sigma}$ [1/m]
    \item $r_w$: Radius of the primary winding wire [m]
    \item $\delta_w$: Skin depth in the winding wire [m]
    \item $\xi$: Dimensionless size factor for wire, $\xi = \sqrt{2} r_w / \delta_w$
    \item $\mu_{app}$: Apparent complex permeability [H/m]
    \item $J_0, J_1$: Bessel functions of the first kind
    \item $\text{ber}, \text{bei}$: Kelvin functions
\end{itemize}

\section{Modeled Ring}

\subsection{3D Representation of the Toroidal Ring}
\begin{figure}[h!]
\centering
\includegraphics[width=0.5\textwidth]{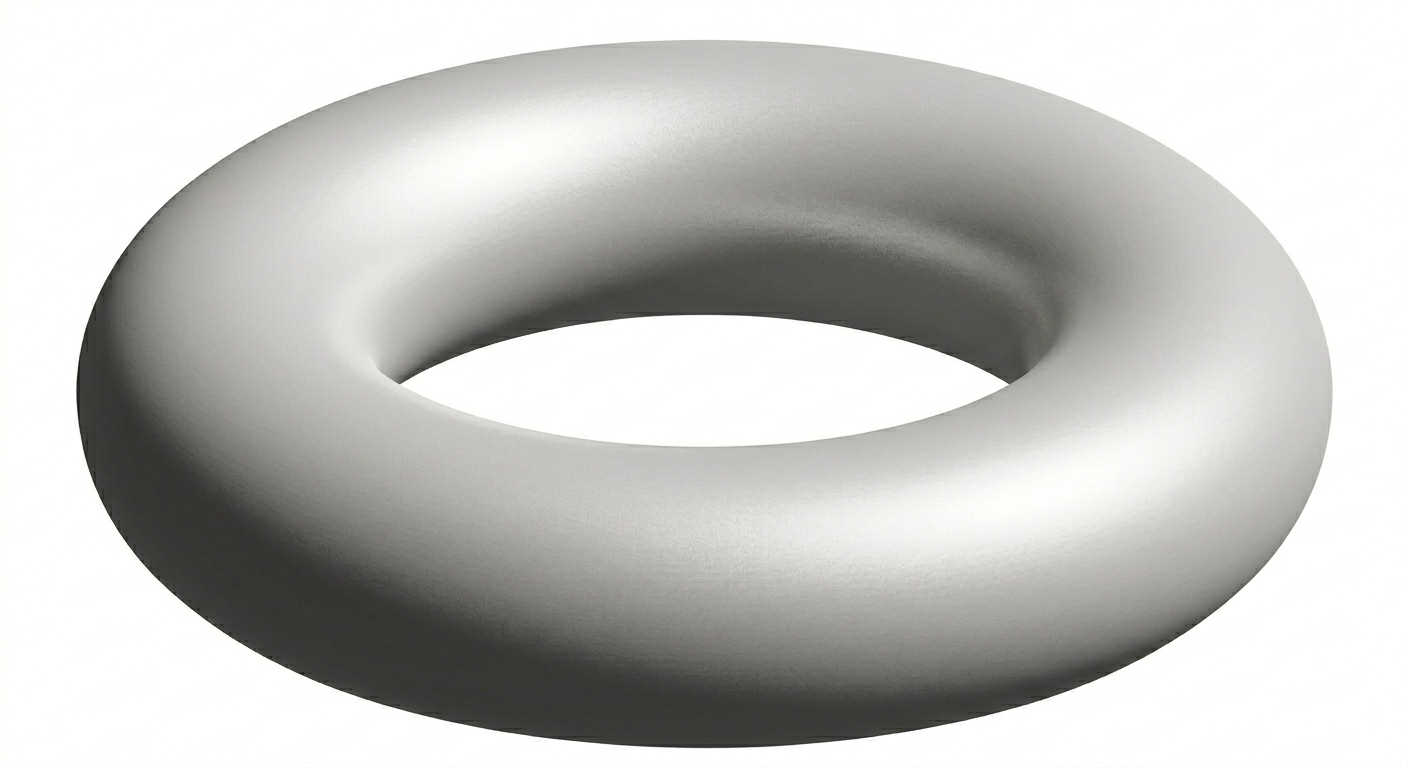}
\caption{Visualization of the Toroidal Ring (Circular Cross-Section)}
\label{ring3D}
\end{figure}

Fig. \ref{ring3D} represents a 3D visualization of the studied ring, a toroidal ring with a circular cross-section.

\begin{figure}[htbp]
    \centering
    \begin{subfigure}{0.4\textwidth}
        \centering
        \includegraphics[width=\textwidth]{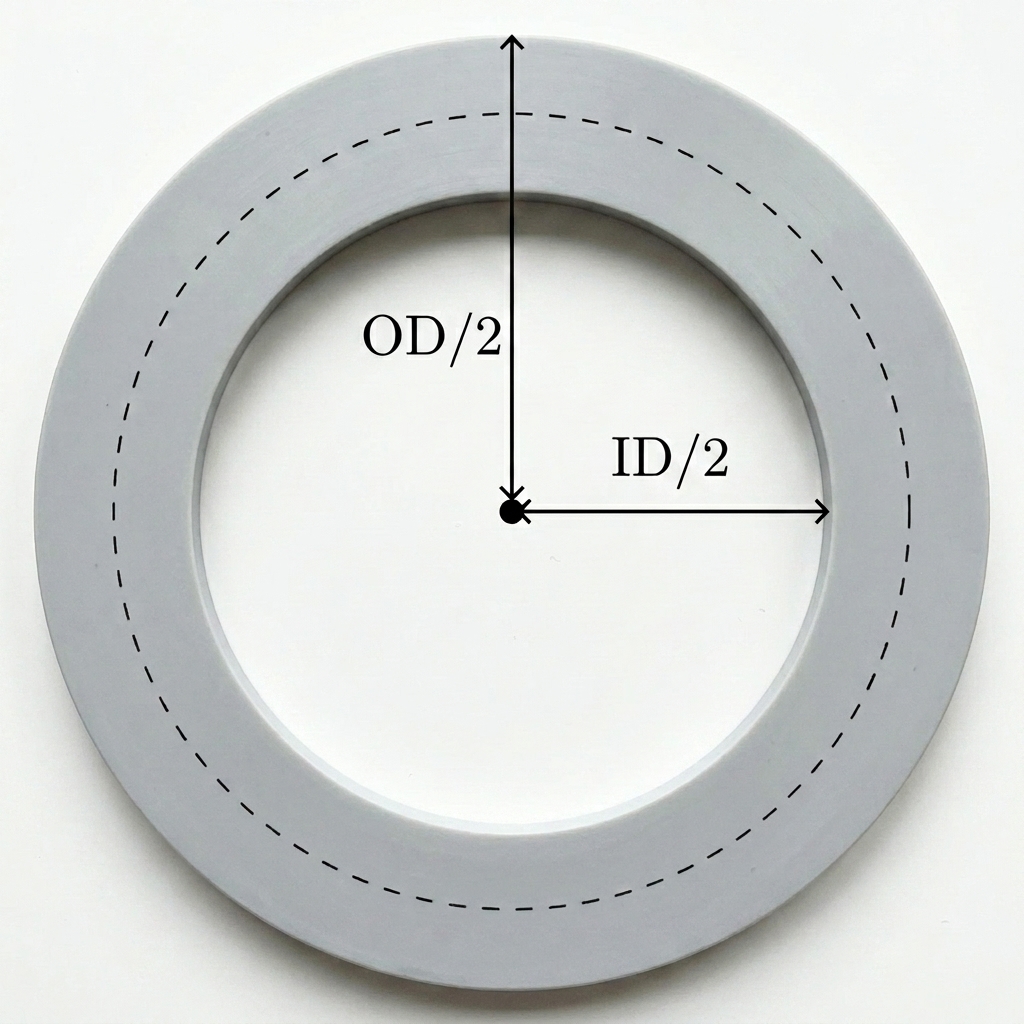}
        \caption{}
        \label{fig:top}
    \end{subfigure}
    \hfill
    \begin{subfigure}{0.45\textwidth}
        \centering
        \includegraphics[width=\textwidth]{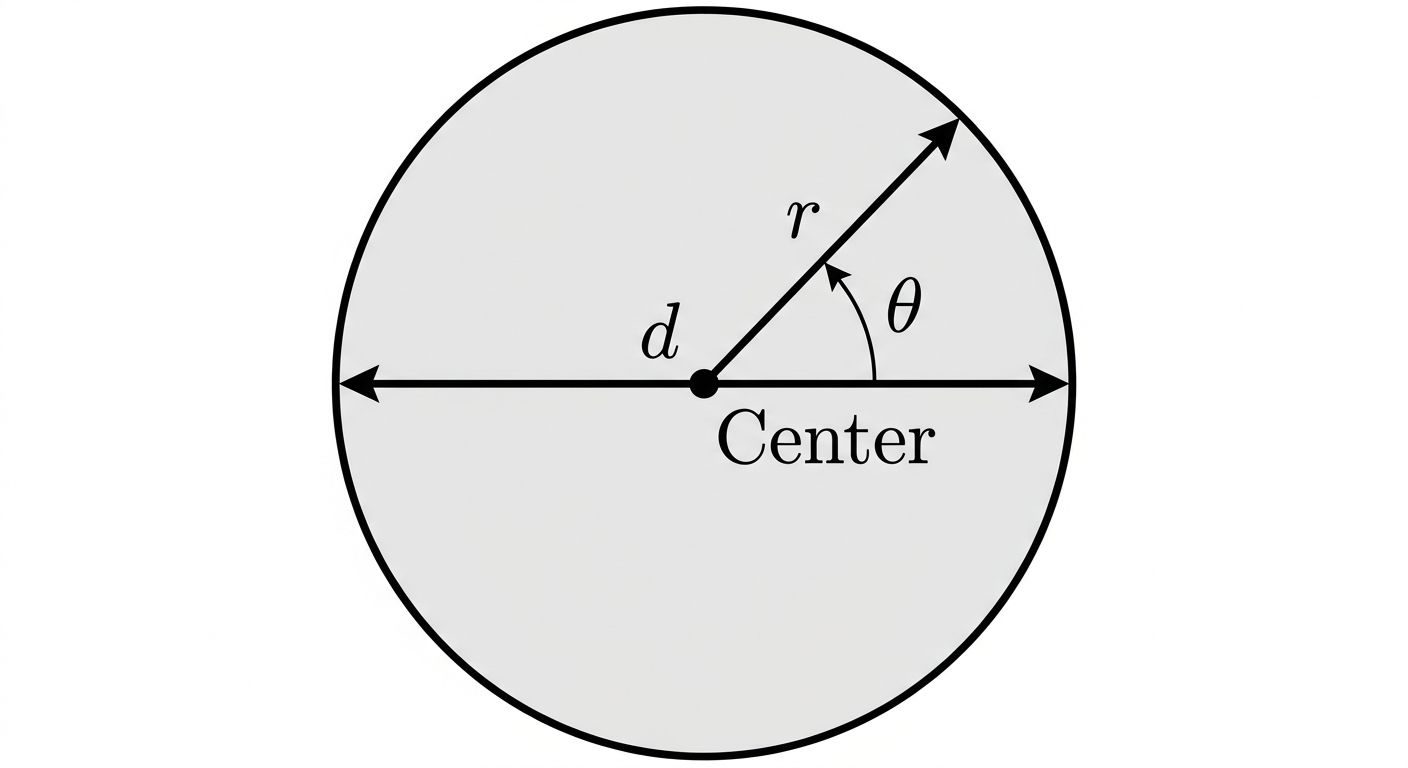}
        \caption{}
        \label{fig:cross}
    \end{subfigure}
    \caption{Toroidal Ring: (a) Top View, (b) Cross-section}
    \label{fig:geometry}
\end{figure}

\subsection{Winding Configuration (Brockhaus Standard)}
\begin{figure}[h!]
\centering
\includegraphics[width=0.5\textwidth]{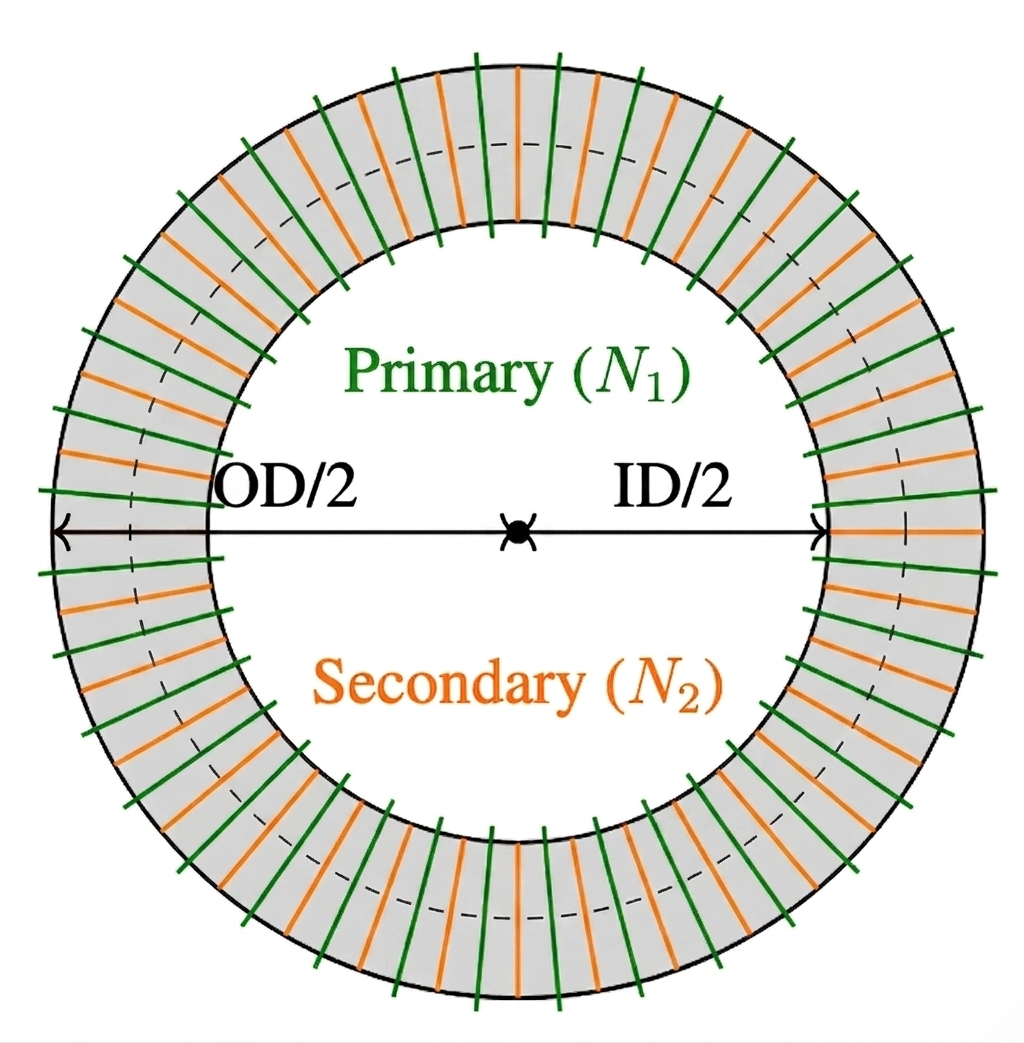}
\caption{Winding Configuration for Circular Core}
\label{ringwinding}
\end{figure}

The dimensions of the ring are illustrated in Fig.~\ref{fig:geometry}. In standard material testing, the secondary winding ($  N_2  $) is wound directly onto the core to accurately capture the true flux linkage, with leakage minimized to negligible levels. The primary winding ($  N_1  $) is then wound uniformly over the secondary. The winding configuration is shown in Fig.~\ref{ringwinding}.

\subsection{Equivalent Circuit Model}
\begin{figure}[h!]
    \centering
    \includegraphics[width=0.99\linewidth]{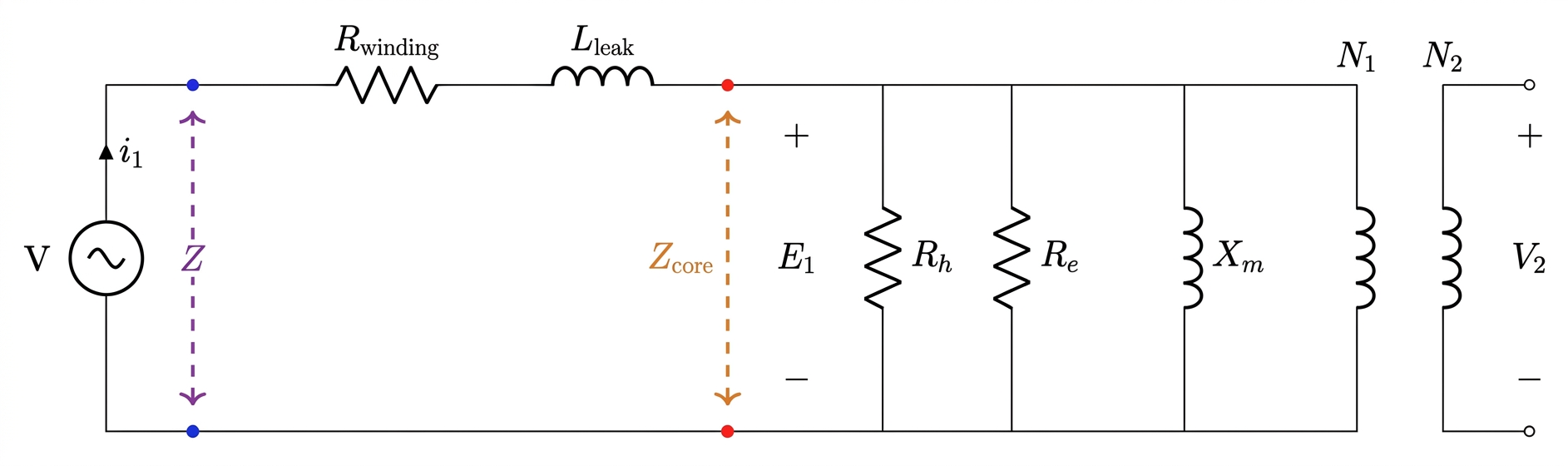}
    \caption{Equivalent magnetic circuit of the wounded ring}
    \label{fig:magcircuit}
\end{figure}

The electrical behavior of the system is illustrated in Fig.~\ref{fig:magcircuit}. The primary voltage $  V_1  $ comprises the voltage drop across the winding resistance, the voltage across the residual leakage inductance $  L_{leak}  $, and the induced voltage $  E_1  $ at the core level. The non-ideal behaviors of the core are modeled by parallel branches representing hysteresis losses ($  R_h  $), eddy current losses ($  R_e  $), and the magnetizing reactance ($  X_m  $).

The transformer circuit is governed by the following equation:
\begin{align}
    V_1 &= R_{winding} i_1 + L_{leak} \frac{di_1}{dt} + E_1 \\
    \frac{E_1}{N_1} &= \frac{V_2}{N_2}
\end{align}

\section{Modeling and Mathematical Proof}

\subsection{Fundamental Maxwell's Equations}
The magnetic core is governed by the generalized Maxwell's equations:
\begin{align}
    \nabla \cdot \mathbf{D} &= \rho_v \\
    \nabla \cdot \mathbf{B} &= 0 \\
    \nabla \times \mathbf{E} &= -\frac{\partial \mathbf{B}}{\partial t}  \\
    \nabla \times \mathbf{H} &= \mathbf{J} + \frac{\partial \mathbf{D}}{\partial t} 
\end{align}
In this model, the quasi-static approximation holds strongly for the operating frequencies under consideration, so the displacement current density ($\frac{\partial \mathbf{D}}{\partial t}$) is neglected. Consequently, the analysis relies primarily on Faraday's law and the simplified Ampère's law ($\nabla \times \mathbf{H} = \mathbf{J}$). All calculations in this paper are performed under no-load conditions.

\subsection{Assumptions and Operational Constraints}
The analytical framework operates under the following key constraints:
\begin{itemize}
    \item $OD/ID \le 1.2$: Ensures a relatively uniform magnetic field $\mathbf{H}$ at the surface across the radius at low frequencies.
    \item $D_m > 10d$: Validates the 1-dimensional radial approximation within the 2D cross-section.
    \item The core material has a uniform complex permeability $\mu_c = \mu' - j\mu'' = |\mu_c| e^{-j\theta}$ and a constant electrical conductivity ($\sigma$).
    \item Operation within the Rayleigh region ensures linearity, avoiding spatial dependencies of $\mu_c$ caused by magnetic saturation.
    \item Zero current at the secondary winding.
\end{itemize}

\subsection{Maxwell's Equations in 2D Plane}
We establish a local polar coordinate system $(\rho, \phi, z)$ within the cross-section of the toroid. The origin $\rho=0$ is the center of the circular cross-section. The $z$-axis of this local system corresponds to the tangential direction (i.e. the mean circumference) of the toroid.

Due to symmetry, the magnetic field is purely axial (in the local $z$-direction): $\mathbf{H} = H_z(\rho) \hat{\mathbf{z}}$, and the induced electric field is purely azimuthal: $\mathbf{E} = E_\phi(\rho) \hat{\boldsymbol{\phi}}$. Considering time-harmonic fields ($e^{j\omega t}$), Maxwell's equations in the core are expressed as:
\begin{align}
    \nabla \times \mathbf{H} &= \sigma \mathbf{E} \label{eq:ampere} \\
    \nabla \times \mathbf{E} &= -j\omega \mathbf{B} = -j\omega \mu_c \mathbf{H} \label{eq:faraday}
\end{align}

Taking the curl of Eq. \ref{eq:ampere} and substituting in Eq. \ref{eq:faraday}:
\begin{equation}
    \nabla \times (\nabla \times \mathbf{H}) = \sigma (\nabla \times \mathbf{E}) = -j\omega \mu_c \sigma \mathbf{H}
\end{equation}
Using the vector identity $\nabla \times (\nabla \times \mathbf{H}) = \nabla(\nabla \cdot \mathbf{H}) - \nabla^2 \mathbf{H}$ and knowing $\nabla \cdot \mathbf{H} = 0$, we obtain the following equation (diffusion equation):
\begin{equation}
    \nabla^2 \mathbf{H} - j\omega \mu_c \sigma \mathbf{H} = 0
\end{equation}

In cylindrical coordinates, assuming a 1D dependency strictly on $\rho$, this simplifies to:
\begin{equation}
    \frac{d^2 H_z}{d\rho^2} + \frac{1}{\rho} \frac{d H_z}{d\rho} - j\omega \mu_c \sigma H_z = 0
\end{equation}
By defining
\begin{equation}
    k^2 = -j\omega \mu_c \sigma
\end{equation}

The equation becomes a standard Bessel differential equation of order zero:
\begin{equation} \label{eq:bessel_H}
    \frac{d^2 H_z}{d\rho^2} + \frac{1}{\rho} \frac{d H_z}{d\rho} + k^2 H_z = 0
\end{equation}

\subsection{Expressions for B and H}
The general mathematical solution to Eq. \ref{eq:bessel_H} is $H_z(\rho) = C_1 J_0(k\rho) + C_2 Y_0(k\rho)$. Since the magnetic field must remain finite at the center of the cross-section ($\rho=0$) and the Bessel function of the second kind $Y_0(0)$ diverges, we set $C_2 = 0$. 

Applying Ampère's law along the mean magnetic path $l_m = \pi D_m$, the field at the boundary surface of the cross-section ($\rho = a$) is constrained by the primary excitation current:
\begin{equation}
    H_{surf} = H_z(a) = \frac{N_1 I}{l_m}
\end{equation}
Thus, the constant is evaluated as $C_1 = \frac{H_{surf}}{J_0(ka)}$, yielding the exact analytical expressions for the internal fields:
\begin{equation}
    \boxed{H_z(\rho) = \frac{N_1 I}{l_m} \frac{J_0(k\rho)}{J_0(ka)}}
\end{equation}
\begin{equation}
    \boxed{B_z(\rho) = \mu_c \frac{N_1 I}{l_m} \frac{J_0(k\rho)}{J_0(ka)}}
\end{equation}

\subsection{Flux Expression}
The total magnetic flux $\Phi$ passing through the circular cross-section is the area integral of $B_z(\rho)$:
\begin{equation}
    \Phi = \int_0^a B_z(\rho) 2\pi \rho d\rho = 2\pi \mu_c \frac{H_{surf}}{J_0(ka)} \int_0^a \rho J_0(k\rho) d\rho
\end{equation}
Applying the integral identity $\int x J_0(x) dx = x J_1(x)$, we deduce:
\begin{equation}
    \Phi = 2\pi \mu_c \frac{H_{surf}}{J_0(ka)} \left( \frac{a}{k} J_1(ka) \right) = \mu_c H_{surf} (\pi a^2) \frac{2 J_1(ka)}{ka J_0(ka)}
\end{equation}
Substituting the cross-sectional area $A = \pi a^2$ and $H_{surf} = N_1 I / l_m$:
\begin{equation}
    \boxed{\Phi = \mu_c \frac{N_1 I A}{l_m} \frac{2 J_1(ka)}{ka J_0(ka)}}
\end{equation}

\subsection{Primary Complex Impedance Z}
The induced electromotive force (EMF) in the primary is $E_1 = j\omega N_1 \Phi$. The impedance of the core $Z_{core} = E_1 / I$ becomes:
\begin{equation}
    \boxed{Z_{core} = j\omega \frac{N_1^2 A}{l_m} \mu_c \frac{2 J_1(ka)}{ka J_0(ka)}}
\end{equation}

Simultaneously, the primary winding AC resistance $R_{winding}$ for circular wires is defined using Kelvin functions \cite{tahaACloss} to strictly account for the skin effect in the copper conductors:
\begin{equation}
    R_{winding} = R_{DC} \frac{\xi}{2} \left[ \frac{\text{ber}(\xi) \text{bei}'(\xi) - \text{bei}(\xi) \text{ber}'(\xi)}{\text{ber}'^2(\xi) + \text{bei}'^2(\xi)} \right]
\end{equation}

Adding the terms, the total complex impedance is:
\begin{equation}
    \boxed{Z = R_{winding} + j\omega L_{leak} + j\omega \frac{N_1^2 A}{l_m} \mu_c \frac{2 J_1(ka)}{ka J_0(ka)}}
\end{equation}

\subsection{Losses Expression}
The different losses can be expressed separately: Winding losses, Eddy Current losses, and Hysteresis losses.

\textbf{1. Winding Loss:}
\begin{equation}
    \boxed{P_w = I_{rms}^2 R_{winding}}
\end{equation}

\textbf{2. Core Losses Separation (Volume Integration):} 
By integrating the Poynting vector over the volume of the toroid, the total power dissipated in the material separates strictly into ohmic heating (eddy currents) and magnetic domain friction (hysteresis):
\begin{equation}
    P_{core} = \underbrace{\int_{Vol} \sigma |E_\phi|^2 dV}_{\text{Eddy Loss } P_e} + \underbrace{\int_{Vol} \omega \mu'' |H_z|^2 dV}_{\text{Hysteresis Loss } P_h}
\end{equation}
From Ampère's Law ($\mathbf{E} = \frac{1}{\sigma} \nabla \times \mathbf{H}$), the induced electric field is $E_\phi(\rho) = \frac{k H_{surf}}{\sigma} \frac{J_1(k\rho)}{J_0(ka)}$. Integrating over the geometry yields the exact loss formulas:
\begin{equation}
    \boxed{P_e = \frac{2\pi l_m |k|^2 |H_{surf}|^2}{\sigma |J_0(ka)|^2} \int_0^a \rho |J_1(k\rho)|^2 d\rho}
\end{equation}
\begin{equation}
    \boxed{P_h = \frac{2\pi l_m \omega \mu'' |H_{surf}|^2}{|J_0(ka)|^2} \int_0^a \rho |J_0(k\rho)|^2 d\rho}
\end{equation}

\textbf{Alternative Circuit Admittance Method:}
Equivalently, one can separate the losses macroscopically using the complex admittance $Y_{core} = 1/Z_{core} = G_{core} - jB_{core}$. The hysteresis conductance $G_h$ is isolated by taking the limit as conductivity approaches zero ($\sigma \to 0$):
\begin{equation}
    G_h = \text{Re}\left( \frac{1}{j\omega \mu_c \frac{N_1^2 A}{l_m}} \right)
\end{equation}
This yields the circuit-based loss representation:
\begin{equation}
    \boxed{P_h = |E_{1,rms}|^2 G_h} \quad \text{and} \quad \boxed{P_e = |E_{1,rms}|^2 (G_{core} - G_h)}
\end{equation}

\subsection{Apparent Permeability}
To separate the effect of eddy currents on the core’s ability to store magnetic energy, we define an apparent permeability $\mu_{app}$. This quantity is extracted from the reactive part of the electrical circuit.

Consider an idealised core in which macroscopic eddy currents are absent. Its inductive behaviour is described solely by a real apparent permeability $\mu_{app}$, leading to a purely inductive impedance $Z_{ideal} = j\omega L_{app}$. With cross-sectional area $A = \pi a^2$, the imaginary part of this impedance is
\begin{equation}
    \text{Im}(Z_{ideal}) = \omega \frac{N_1^2 A}{l_m} \mu_{app}
\end{equation}

In contrast, the exact analytical impedance of a real core, accounting for skin effect, eddy currents, and complex permeability, is expressed as:
\begin{equation}
    Z_{core} = j\omega \frac{N_1^2 A}{l_m} \mu_c \frac{2 J_1(ka)}{ka J_0(ka)}
\end{equation}

Using the definition of the wave number $k^2 = -j\omega \mu_c \sigma$, we can replace $j\omega \mu_c$ with $-k^2/\sigma$. This transforms the core impedance into
\begin{equation}
    Z_{core} = -\frac{N_1^2 A k}{\sigma a l_m} \frac{2 J_1(ka)}{J_0(ka)}
\end{equation}

The actual inductive reactance of the physical core is given by the imaginary part of this complex expression:
\begin{equation}
    \text{Im}(Z_{core}) = \text{Im} \left( -\frac{N_1^2 A k}{\sigma a l_m} \frac{2 J_1(ka)}{J_0(ka)} \right)
\end{equation}

We now require that the equivalent idealised model reproduces the true inductive behaviour. Equating the two imaginary parts yields
\begin{equation}
    \omega \frac{N_1^2 A}{l_m} \mu_{app} = \text{Im} \left( -\frac{N_1^2 A k}{\sigma a l_m} \frac{2 J_1(ka)}{J_0(ka)} \right)
\end{equation}

This leads to:
\begin{equation}
    \boxed{\mu_{app} = \text{Im} \left( -\frac{k}{\omega \sigma a} \frac{2 J_1(ka)}{J_0(ka)} \right)}
\end{equation}

\section{Summary of Main Formulas}
The main formulations for the circular cross-section toroid are summarized in Table \ref{tab:summary}.

\renewcommand{\arraystretch}{1.5}
\begin{table}[h]
\centering
\caption{Summary of Main Formulations}
\label{tab:summary}
\begin{tabular}{p{5.5cm} p{10cm}}
\toprule
\textbf{Quantity} & \textbf{Expression} \\
\midrule
Magnetic field $H_z(\rho)$ & $\displaystyle H_z(\rho) = \frac{N_1 I}{l_m} \frac{J_0(k\rho)}{J_0(ka)}$ \\[1em]
Flux density $B_z(\rho)$ & $\displaystyle B_z(\rho) = \mu_c \frac{N_1 I}{l_m} \frac{J_0(k\rho)}{J_0(ka)}$ \\[1em]
Magnetic flux $\Phi$ & $\displaystyle \Phi = \mu_c \frac{N_1 I A}{l_m} \frac{2 J_1(ka)}{ka J_0(ka)}$ \\[1em]
Apparent permeability $\mu_{app}$ & $\displaystyle \mu_{app} = \frac{l_m}{\omega N_1^2 A} \text{Im} \left( Z_{core} \right)$ \\[1em]
Core impedance $Z_{core}$ & $\displaystyle Z_{core} = j\omega \frac{N_1^2 A}{l_m} \mu_c \frac{2 J_1(ka)}{ka J_0(ka)}$ \\[1em]
Eddy current loss $P_e$ & $\displaystyle P_e = |E_{1,rms}|^2 (G_{core} - G_h)$ \\[1em]
Hysteresis loss $P_h$ & $\displaystyle P_h = |E_{1,rms}|^2 G_h$ \\[1em]
Winding AC resistance $R_{winding}$ & $\displaystyle R_{winding} = R_{DC} \frac{\xi}{2} \left[ \frac{\text{ber}(\xi) \text{bei}'(\xi) - \text{bei}(\xi) \text{ber}'(\xi)}{\text{ber}'^2(\xi) + \text{bei}'^2(\xi)} \right]$ \\[1em]
\bottomrule
\end{tabular}
\end{table}

\section{Results and Analysis}

The analytical models derived in the previous sections are employed to assess the electromagnetic behaviour of the toroidal ring with a circular cross-section. Table~\ref{tab:sim_params_circ} lists the geometric, material, and excitation parameters used in this assessment. The frequency response was computed over a sweep from 10~Hz to 1~MHz with 500 discrete sampling points.
\renewcommand{\arraystretch}{1}
\begin{table}[htbp]
    \centering
    \caption{Material and Ring Parameters}
    \label{tab:sim_params_circ}
    \begin{tabular}{lcc}
        \toprule
        \textbf{Parameter} & \textbf{Symbol} & \textbf{Value} \\
        \midrule
        \textbf{Geometry} & & \\
        Radius of circular cross-section & $a$ & 2.5 mm \\
        Mean diameter of the ring & $D_m$ & 50 mm \\
        Number of primary turns & $N_1$ & 60 \\
        \midrule
        \textbf{Material Properties} & & \\
        Real relative permeability & $\mu_r'$ & 500 \\
        Hysteresis loss angle & $\theta$ & 0.1 rad \\
        Electrical conductivity & $\sigma$ & 900 S/m \\
        \midrule
        \textbf{Excitation} & & \\
        RMS primary current & $I_{rms}$ & 1.0 A \\
        Frequency range & $f$ & $10^1$ -- $10^6$ Hz \\
        \bottomrule
    \end{tabular}
\end{table}

\subsection{Spatial Distribution of Magnetic Flux Density}

\begin{figure}[htbp]
    \centering
    \begin{subfigure}[b]{0.45\textwidth}
        \centering
        \includegraphics[width=\textwidth]{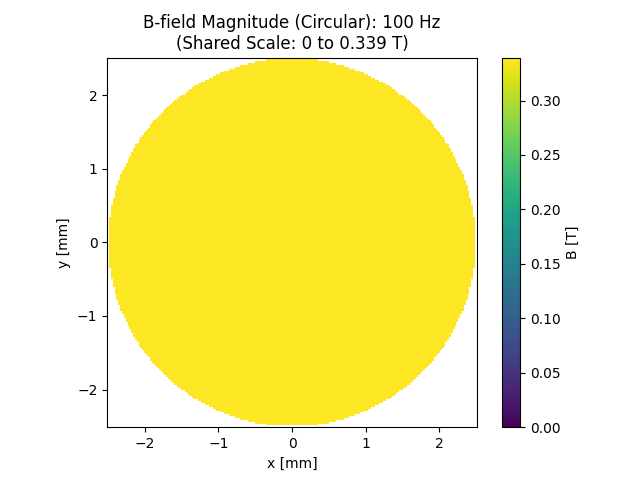}
        \caption{100 Hz}
        \label{fig:B100_circ}
    \end{subfigure}
    \hfill
    \begin{subfigure}[b]{0.45\textwidth}
        \centering
        \includegraphics[width=\textwidth]{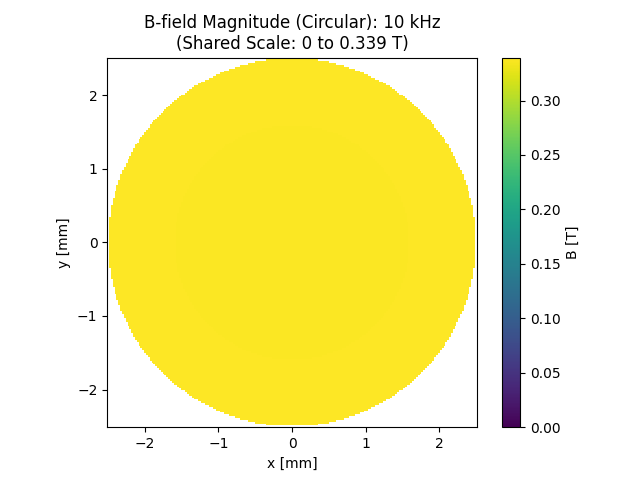}
        \caption{10 kHz}
        \label{fig:B10k_circ}
    \end{subfigure}
    
    \vspace{0.5cm}
    
    \begin{subfigure}[b]{0.45\textwidth}
        \centering
        \includegraphics[width=\textwidth]{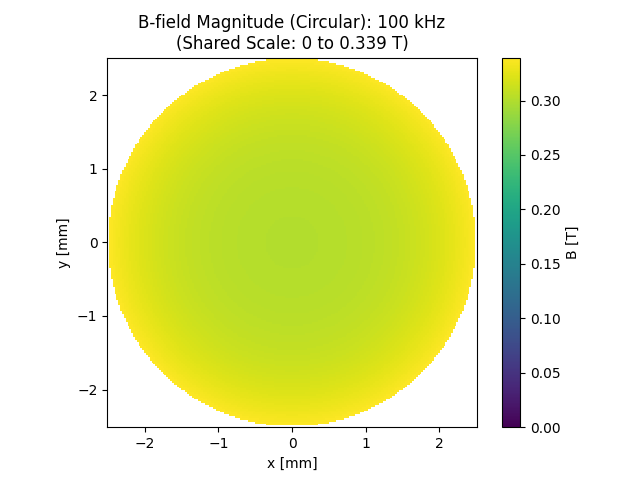}
        \caption{100 kHz}
        \label{fig:B100k_circ}
    \end{subfigure}
    \hfill
    \begin{subfigure}[b]{0.45\textwidth}
        \centering
        \includegraphics[width=\textwidth]{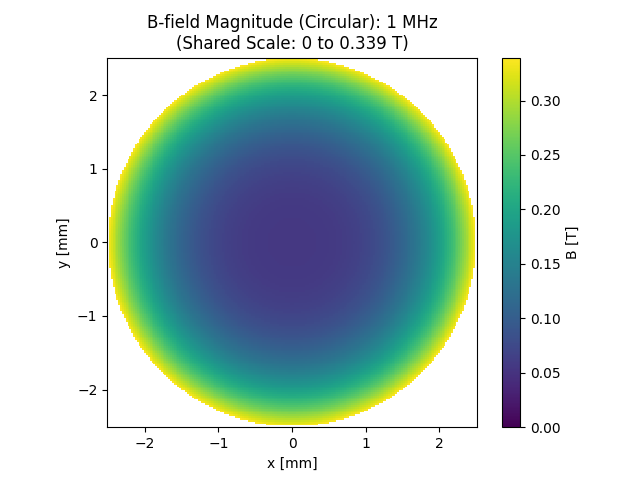}
        \caption{1 MHz}
        \label{fig:B1M_circ}
    \end{subfigure}
    \caption{Magnetic Flux Density magnitude across the circular cross-section at 4 different frequencies.}
    \label{fig:B_profiles_circ}
\end{figure}

The two-dimensional spatial variation of the magnetic flux density magnitude $|B_z(\rho)|$ across the circular cross-section is shown in Figure~\ref{fig:B_profiles_circ} for four different frequencies.

At low frequency (100~Hz), the electromagnetic skin depth is much larger than the radius $a$, leading to a nearly uniform flux density of approximately 0.339~T. When the excitation frequency reaches 100~kHz, the induced eddy currents produce a counter-field that starts to shield the core's centre. At the upper limit of 1~MHz, the skin effect becomes severe: the magnetic flux is almost entirely expelled from the central region, resulting in a hollow distribution where the flux remains confined to a thin skin layer along the periphery.

\subsection{Apparent Permeability and Impedance Characteristics}

\begin{figure}[ht]
    \centering
    \includegraphics[width=0.65\textwidth]{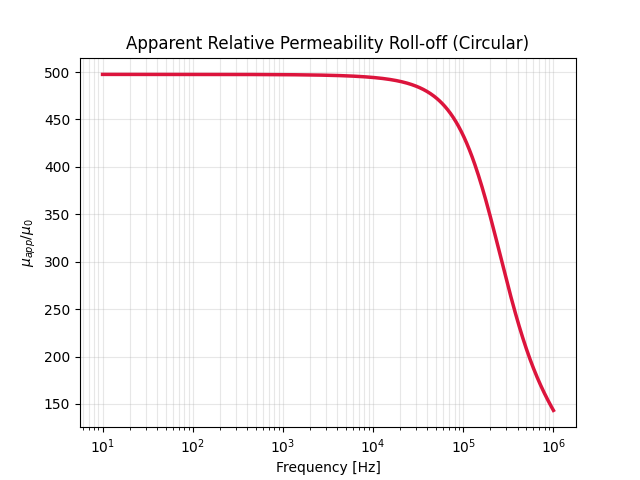}
    \caption{Apparent Relative Permeability vs Frequency}
    \label{fig:mu_app_circ}
\end{figure}

The macroscopic effect of this shielding is captured by the apparent relative permeability $\mu_{app}/\mu_0$ (Fig.~\ref{fig:mu_app_circ}). Up to about 20~kHz, the apparent permeability stays constant at its intrinsic value of 500. Above this corner frequency, the core's effective capacity to sustain magnetic flux declines rapidly, falling to approximately 140 at 1~MHz.

\begin{figure}[ht]
    \centering
    \includegraphics[width=0.65\textwidth]{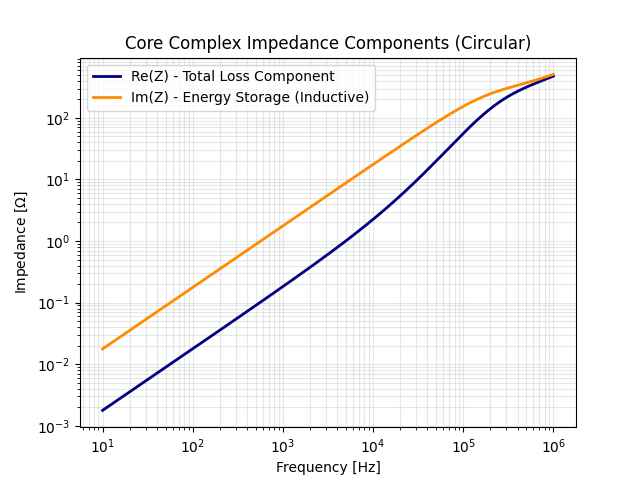}
    \caption{Core complex impedance components}
    \label{fig:impedance_circ}
\end{figure}

This transition is reflected in the primary complex impedance components (Fig.~\ref{fig:impedance_circ}). At low frequencies, the inductive reactance $\text{Im}(Z)$ increases linearly, but it begins to saturate as the apparent permeability rolls off. Meanwhile, the resistive component $\text{Re}(Z)$, representing the total core losses, rises with frequency. Significantly, at high frequencies, the loss component $\text{Re}(Z)$ approaches the inductive component $\text{Im}(Z)$ in magnitude, marking a shift toward a more dissipative behaviour.

\subsection{Loss Separation and Attenuation Mechanisms}

\begin{figure}[ht]
    \centering
    \includegraphics[width=0.65\textwidth]{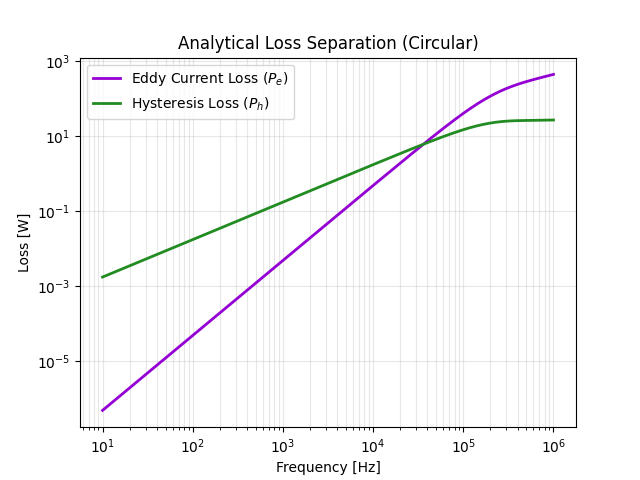}
    \caption{Hysteresis and Eddy current losses}
    \label{fig:losses_circ}
\end{figure}

Figure~\ref{fig:losses_circ} presents the analytical decomposition of losses. At low frequencies, hysteresis loss ($P_h$) dominates, increasing linearly with frequency. In the quasi-static regime, eddy current loss ($P_e$) grows with the square of frequency ($\omega^2$) and eventually surpasses hysteresis loss around 40~kHz.

As in the square cross-section case~\cite{TahaArxivRingS}, the growth rates of both loss components diminish at high frequencies. This behaviour is attributed to magnetic shielding: once the skin effect becomes dominant, the total magnetic flux $\Phi$ inside the core decreases. Because eddy current losses are driven by the rate of change of flux, the substantial reduction in internal flux density restricts the available volume. Consequently, the power loss curves exhibit the behaviour observed in the MHz range. These results are obtained under constant-current excitation ($I_{rms} = 1$~A). Under constant-flux density excitation, the losses would increase much more rapidly — a situation that occurs, for instance, in constant-flux measurements such as those performed with Brockhaus or Iwatsu equipment.
\clearpage
\bibliographystyle{ieeetr}
\bibliography{references}

\end{document}